# OPTICAL DIFFRACTION-TRANSITION RADIATION INTERFEROMETRY AND ITS APPLICATION TO THE MEASUREMENT OF BEAM DIVERGENCE

A. G. Shkvarunets[a], R. B. Fiorito[a,b,*] and P. G. O'Shea[a,c]

[a]Institute for Research in Electronics and Applied Physics, [c]Department of Computer and Electrical Engineering, University of Maryland, College Park, MD 20742

[b] TR Research Inc., 3 Lauer Terrace, Silver Spring, MD 20901

*Abstract*

Optical transition radiation interferometry (OTRI) has been shown to be a very useful technique to measure the divergence of electron beams with energies in the range of 15-100 MeV. However, application of this method to low energy or very high quality beams is limited by scattering in the front foil of the interferometer. To overcome this limitation we propose to use a perforated front foil. For the beam energy and hole sizes we are considering, the unscattered beam electrons passing through the holes will produce diffraction radiation (ODR). The total radiation produced from the first and second foils then will be a spatially coherent sum of ODR and OTR from unscattered and scattered electrons. By controlling the number and size of the perforations, the inter-foil spacing, the thickness of the first foil and the wavelength and band pass of the observed radiation, the coherent interferences due to the unscattered portion of the beam can be isolated and observed. The visibility of these interferences can then be used to determine the rms beam divergence. We have developed a general computer code which can be used to calculate diffraction radiation from any type of perforation and another code to compute the ODR-OTR interference pattern for a given set of beam and optical parameters. These codes are employed in the design of an interferometer to measure the divergence of the ATF accelerator operating at 30 MeV. This beam will be used in an initial proof of principle experiment for the ODR-OTR interferometer. We present the results of our code calculations which show that the expected divergence, 200 micro radians, can be easily measured.

---

[*] Corresponding author. Tel. and fax: (301) 587 7575; Email address: rfiorito@trresearch.com

# 1 INTRODUCTION

A typical optical transition radiation (OTR) interferometer consists of two parallel thin foils, oriented at 45 degrees with respect to the electron beam. The beam produces forward directed OTR from the first foil and backward directed OTR from the second. When the distance between the foils is comparable to the vacuum coherence length $L_V \sim \gamma^2\lambda$, interferences between these two sources are observed; the visibility of the interference pattern is a function of the rms beam divergence [1].

Scattering in the first foil of the interferometer limits this method when the beam divergence is less than about $0.05\gamma^{-1}$ or when scattering in the first foil dominates the divergence. Practically speaking, the latter occurs when the beam energy is lower than about 10 MeV. However, even when the beam energy is larger, if the beam quality is very high, the foil scattering may be sufficient to mask the real beam divergence. To overcome this limitation, we plan to use a perforated first foil as shown in Figure 1. The hole size and spacing are chosen to be much smaller than the rms beam size. In addition, the holes are sufficiently small and numerous so that significant ODR will be produced. The conditions for significant DR production are (a) the characteristic hole dimensions $a \leq \gamma\lambda/2\pi$, where $\gamma$ is the Lorentz factor of the beam and $\lambda$ is the wavelength of the observed DR and (b) the transparency of the perforated foil is about 50%.

The total output light intensity distribution observed from such a device will be the coherent sum of forward directed ODR and OTR from the perforated foil and backward directed OTR from the second foil. There will be two contributions to the interference pattern ODR-OTR radiation: one from the unscattered portion and the second from the scattered portion of the beam. By proper the choice of front foil thickness, inter-foil spacing, hole size, number of holes, spacing of the holes and the optical band pass, the interference fringes produced by the unscattered electrons can be made clearly visible over the background radiation produced by the scattered portion of beam. A sufficiently large scattering in the front foil will destroy the visibility of interferences produced by the scattered electrons, rendering a smooth background radiation pattern. The amount of scattering can be controlled by choosing the thickness and material density of the front foil for the particular beam energy and the range of beam divergences under investigation. The rms divergence of the *unperturbed* beam can then be determined by an analysis of the observed interference pattern riding above the scattered background.

Also, as with conventional OTR interferometry [1], the orthogonal (x, y) components of the divergence can be separately obtained by observing the interferences with a polarizer and adjusting the beam focus to either an x or y waist condition. In addition to the rms divergences, the orthogonal rms emittances of a non space charged dominated beam can be determined by also measuring the rms beam waist sizes. This can be easily done experimentally by imaging the beam in OTR at the position of the second foil.

## 2 APPROACH

In order to design the interferometer to be sensitive to a given beam divergence we have developed two computer codes. The code CONVOLUTION convolves the total intensity (see Eq. (3) below) with the transmission function of the optical pass band filter and beam angular distributions. A second code BEAM DR is used to calculate the individual fields and intensities of ODR and OTR, which are inputs to code CONVOLUTION.

BEAM DR is extremely general and powerful. It is capable of calculating DR from any type of hole, for any position of the electron passing through the hole and for any distribution of holes in a perforated foil. In addition, the code can calculate the distribution of DR at any distance from the radiating source.

### 2.1 ODR - OTR Interferences – Code CONVOLUTION

In an OTR interferometer the electron beam passes two continuous metallic foils. The beam produces similar intense OTR from each foil. If the first foil is perforated, the radiation from the first foil is a mixture of ODR and OTR. In either case, when the second foil a highly conductive metallic mirror, the reflectivity of the second foil is essentially unity in the visible part of the spectrum and the combined intensity of the radiation from the first (1) and second (2) foils can be written as:

$$I(\mathbf{q},\lambda) = I_1(\mathbf{q}) + I_2(\mathbf{q}) - E_1(\mathbf{q})E_2(\mathbf{q})\cos\Psi(\mathbf{q},\mathbf{q}_e,\lambda) \quad (1)$$

where $\mathbf{q}$ is the observation angle, $\lambda$ is the wavelength of the observed radiation, $I_1(\mathbf{q}) = 1/2\, E_1^2(\mathbf{q})$, $I_2(\mathbf{q}) = 1/2\, E_2^2(\mathbf{q})$, and $E_1(\mathbf{q})$ and $E_2(\mathbf{q})$ are ODR and OTR intensities and fields produced by foils (1) and (2), respectively; $\psi$, the phase shift between forward ODR produced from the rear surface of foil (1) and backward OTR produced from the front surface of foil (2) is given by:

$$\Psi = (2\pi d/\lambda)\left(\beta^{-1}\cos^{-1}\mathbf{q}_e - \cos(\mathbf{q}-\mathbf{q}_e)/\cos\mathbf{q}_e\right) \quad (2)$$

where $d$ is the distance between the foils, $\beta = V/c$, $V$ is the electron velocity, c is the velocity of light and $\mathbf{q}_e$ is the trajectory angle of the electron. When the phase $\Psi = \pi$, the electron trajectory angle $q_e \ll g^{-1}$ and the observation angle $q \sim g^{-1} \ll 1$, the distance $d$ is equal to the familiar vacuum coherence length for TR given by $L_V = (l/2p)/(g^{-2} + q^2)$.

The term $\cos Y$ in Eq. (1) introduces angular interferences (fringes) which are modulated by the single foil intensity functions $I_{1,2}.(q)$. The visibility of these fringes can be very sensitive to any change in the electron trajectory angle $q_e$. In comparison, the single foil intensity, which is a slowly varying function of $q$, is much less sensitive to divergence. Note that fringes (i.e. the function $\cos Y$) also depends on $l$, the wavelength of the observed radiation, whereas the functions $I_{1,2}$, for a highly reflective mirror, are nearly independent of wavelength.

In order to experimentally observe the fringe modulation with good visibility, a narrow band interference filter must be used. To calculate the effect of this filter on the observed intensity, the pass band filter function must be convolved with Eq. (1). The effect of the electron beam divergence on the total intensity can then be calculated by convolving the resultant function with the beam angular distribution, e.g. a Gaussian distribution of beam angles. Then, by comparing the theoretical convolved patterns with measured ones, the rms divergence of the beam can be determined.

A perforated foil with hole sizes and spacing much smaller than the beam size splits beam into two fractions: unscattered and scattered. The forward radiation produced by the unscattered portion of the beam is clearly diffraction radiation. These scattered electrons also produce transition radiation in the solid portion of the foil. However, this component is strongly affected by the presence of neighboring holes and can also be considered as a form of diffraction radiation as well. We will therefore refer to the radiation from *all* the electrons in the perforated foil as DR or ODR in the remainder of this paper. Since the second foil (2) is continuous and large compared to $\gamma\lambda/2\pi$, the radiation from it is normal optical transition radiation (OTR).

The total x or y component of the two foil light intensity can then be written as:

$$I_{Total} = \{I_{1U} + I_{2U} - \cos\Psi \sum_U E_{1e} \cdot E_{2e}\} + \{I_{1S} + I_{2S} - \cos\Psi \sum_S E_{1e} \cdot E_{2e}\} \quad (3)$$

where $I_{1,2,U,S} = \frac{1}{2}\sum_{U,S} E_{1,2e}^2$ are the x or y intensity components from foils (1) or (2) for unscattered (U) or scattered beams (S) and $E_{1,2e}$ are the x or y components of the radiation field for a single electron produced by foils (1) or (2). In this expression the terms $I_{2U}(q)$ and $I_{2S}(q)$ are the OTR angular intensities radiated by beam fractions from the second continuous foil and are independent of the spatial distribution of the beam. However, the ODR intensity $I_{1U}(q)$ and $I_{1S}(q)$ produced by the unscattered and scattered beam fractions from the first foil, and the cross terms $\sum_{U} E_{1e} \cdot E_{2e}$ and $\sum_{S} E_{1e} \cdot E_{2e}$ will also depend on the electron's spatial distribution with respect to the perforations as well as their pattern on the first foil.

## 2.2 Optical Diffraction Radiation – Code BEAM DR

The code BEAM DR calculates the fields and intensities of ODR produced by an electron passing through a perforated foil at any position in the hole. The model which is used in BEAM DR is the wave optics model. This model is attractive because it is naturally applicable to both the perforated holes and surrounding conductive surfaces and allows one to arrive at the solution for the fields at the observation point directly from the sources fields. As a result the code is quite fast and details of the source structure can be taken into account directly without any approximation. This code can naturally calculate the OTR fields and intensities for a continuous foil as well.

In the case of a perfectly conducting target the fields inside the conductor are zero. For simplicity let us consider a plane electromagnetic wave normally incident on conducting surface. The boundary conditions for the electric and magnetic field components $E_{rx}$ and $B_{ry}$ can be written as: $E_x = 0$ and $B_y = J_x$, the induced surface current on the foil which is the source for the emitted radiation. If the target is a flat polished mirror, the surface current can be replaced by an image wave, which is incident on the mirror surface from inside the conductor. The image wave is constructed in such a way that superposition of electric an magnetic fields $E_{rx}$, $B_{ry}$ of the real incident wave and $E_{ix}$, $B_{iy}$ of the image wave on the surface of the mirror satisfies the boundary conditions. In this case the boundary conditions can be written as: $E_{rx} + E_{ix} = 0$ and $B_{ry} = B_{iy}$. The field components of the equivalent image wave are: $E_{ix} = -E_{rx}$ and $B_{iy} = B_{ry}$. The image wave in this case is the mirror image of the real wave. The field of the image wave on the surface of the mirror is considered to be a Huygens source field. The spatial distribution of scattered or secondary radiation can than be calculated by using the Huygens Fresnel Integral:

$$U_{x,y} = \frac{k}{2\pi i} \int_{S_f} \frac{u_{x,y} e^{ikR} \cos q}{R} dS_f \quad (4)$$

where $U_{x,y}$ is the complex field component at an arbitrary observation point, $k=2\pi/\lambda$, $R$ is the distance from $dS_f$ the differential element of area of the source field, to observation point, $S_f$ is the area of the source field distribution and $u_{x,y}$ is the complex field component of the source field [2].

The electron electromagnetic field can be presented as superposition of Fourier waves traveling along electron trajectory. The amplitude of the frequency - longitudinal wave number Fourier components of the free space radial electric and azimuthal magnetic fields of the electron are given by:

$$E_r = e\alpha K_1(\alpha r)/pV \quad (5)$$

$$B_f = \beta E r \quad (6)$$

where V is the electron velocity and $K_1(\alpha r)$ is the Hankel function with imaginary argument and $\alpha = \omega/(V\gamma)$. In the wave optics model the Fourier components of the electron's field are monochromatic azimuthally symmetric plane waves propagating along electron trajectory without divergence (non radiating). If this wave is incident on a metallic mirror it induces a non-zero current on the mirror surface with boundary conditions $E_r = 0$ and $B_j = J_r$. These boundary conditions will be satisfied if the image wave fields are: $E_{ri} = -E_{re}$ and $B_{je} = B_{ji} = J_r$. The image wave in this case is the mirror image of the primary wave provided by the moving field of the electron. This image wave can be considered as a wave produced by a positive image charge traveling as the mirror image of the electron. As the result, backward radiation is produced in the direction of specular reflection of the incident wave from the mirror. For a continuous surface this radiation is the familiar backward reflected transition radiation; in the case of a foil with a hole, it is backward diffraction radiation. In the latter case, the radiation is the reflection of the electron's field from the metallic portion of the foil, and the surface current is induced by the fields of the electron on the conducting surface. For illustration, a sketch of the electron, image and radiation waves is presented in Figure 2. for the case of a continuous surface.

Figure 2 also shows the wave picture for forward TR, which is produced by an electron emerging from the metal foil. In this case the situation is a little more complicated. The boundary conditions in the metal foil can be written as: $E_r = 0$ and $B_j = -J_r$, because now all the fields are zero on the left side of the boundary (i.e. inside the metal). The fields of the image wave which satisfies these conditions are: $E_{ri} = -E_{re}$ and $B_{ji} = -B_{je} = J_r$. The fields of the image wave are the same as those of the primary wave but with reversed sign. The image wave in this case is equivalent to the field of a positive electron moving in the same direction as a real electron inside the metal. Note that the image of the charged particle placed inside the metal is an oppositely charged particle located at the same point. As a result the source of forward TR is the forward directed image wave field traveling along the electron trajectory. Note again that the actual source of the forward TR is the induced surface current.

The total field in free space is superposition of free space electron and TR fields. In the case of forward TR the resulting fields are zero near the foil, and the total electromagnetic field increases along the electron trajectory. The OTR field gradually diverges and the electron's field is reestablished at some distance from the foil, i.e. the vacuum formation or coherence length. In the case of DR the situation is the same for TR except the source currents of DR are confined to the metal portion of the surface, i.e. there is no radiation from the hole. In other terms, the image waves are limited to the portion of screen surrounding the aperture. Thus backward and forward DR are produced by a position electron image wave surrounding the hole.

We developed two versions of the BEAM DR code. Version (1) calculates the angular distribution of diffraction radiation (i.e. the far field or radiation zone solution); version (2) calculates the spatial distribution of DR observed on a flat screen placed at an arbitrary distance from the source. BEAM DR automatically satisfies Babinet principle for complex field amplitude because the solution is the integral over the source surface where non-zero field is attributed only to the metal portion of the foil (i.e. the hole is not a source) [3]. Our code results show that in special cases, for example, DR from a disk or circular hole in a metallic screen when the electron travels directly through the center, or TR from a continuous screen, that the complex field solution can be presented in a simple, real form. Both these cases are represented in Figure 3. where the angular distribution of the radial component of the radiated electric field in real form is shown.

Figure 3. shows the electric field for various conditions. The solid line shows the electric field produced from a continuous metal target (OTR), the dotted line shows the DR field observed from an electron passing through the center of a metal disk with the radius $a = 1/\alpha = \gamma\lambda/2\pi$, and the dashed line is the DR field produced from a hole with the same radius. In accordance with Babinet's Principle, the sum of the fields from the disk and hole add to give the field from the continuous surface (dot-dashed line). The sum overlaps the OTR solid line exactly. We have additionally verified that the code solution for OTR is independent of wavelength, and that the intensity as a function of angle measured in unit of $\gamma^{-1}$ is proportional to $\gamma^2$ in accordance with the well-known properties of OTR.

Figure 4. shows the *x* polarized DR intensity produced by a single edge metallic foil with its edge parallel to the *y* direction. The solid line in Figure 4. corresponds to OTR; the dotted line corresponds to DR radiation produced when the electron is at the edge; the dashed line is DR from an electron located at a distance $l = +0.75/\alpha$ (i.e. inside the foil); and the dot-dashed line is DR

from an electron located at a distance l = - 0.75/α (i.e. outside from the foil). Figure 4. shows that the intensities of the latter two DR intensities at zero angle of observation are identical. The code solution for the fields indicates that the phase shift of the corresponding DR fields is π. This result is another demonstration of Babinet's principle.

BEAM DR also allows us to calculate field and intensity distributions on a flat screen placed at any distance L from the source. In this case the shape of distribution additionally depends on L. To verify the code we compared the code calculations to the analytic results of Verzilov [4]. To do this we calculated the OTR intensity distribution on a flat screen for three values of distance, L=10 $L_R$ , L=$L_R$ and L=0.1 $L_R$, where $L_R$ = γ/α is the OTR analog of the Rayleigh length in wave optics. The results are shown in Figure 5. The calculations, which were done in the same units as those presented in Figure 3. of Ref. [4] are in excellent agreement with analytic results in all cases.

Additional comparisons of BEAM DR calculations with analytic far field DR solutions that are available for a circular aperture [2], a slit [3] and a single edge [5] show excellent agreement as well.

**3 APPLICATION OF CODES TO THE DESIGN OF AN ODR-OTR INTERFEROMETER**

BEAM DR and CONVOLUTION were used to design an interferometer, which is capable of measuring the divergence of the Brookhaven National Laboratory's ATF accelerator operating at 30 MeV. The measured normalized emittance of ATF is approximately 1 micron so that the local divergence for a beam size of 100 microns at 30 MeV is about 200 micro radians [6].

**3.1. BEAM DR Code Results**

Code BEAM DR was used to accurately calculate the DR fields and intensity from a mesh with rectangular perforations. The period of the rectangular holes was 25.8 microns and the holes were 16.8 X 16.8 microns in size. The ratio of the area of holes to the foil area is 0.42. The calculations were done for foil oriented at an angle of 45 degrees with respect to the direction of the electron beam. Figure 6. gives a schematic view of a section of the perforated foil, showing the array of holes and, for illustration, the position and fields of an unscattered electron, which is traveling near the upper left hand corner of one of the holes.

The code calculates the backward radiation fields observed in a plane (X,Y), which is perpendicular to the direction of specular reflection of the incident electron wave field, i.e. at 90 degrees with respect to the velocity of the electrons. The projected angles of observation onto this plane, scaled in units of $γ^{-1}$ , are X = γ$q_x$ and Y= g$q_y$.

Figure 7. shows X scans of ODR and OTR intensities components, polarized parallel to the direction X, from the first and second foils of the unscattered portion of the beam; i.e. $2I_{1U}, 2I_{2U}$ and the first summation term $\sum_U E_{1e} E_{2e}$ in Equation (1).

## 3.2 CONVOLUTION Code Results

The results of CONVOLUTION code are presented in Figure 8. for an observation wavelength $\lambda = 650$ nm, filter pass band $\Delta\lambda/\lambda = 0.03$, foil separation L = 30mm and a beam divergence due to scattering in foil (1), $\sigma_S = 10$ mrads. Under these conditions the scattering destroys the spatial coherence (interference fringes) of the radiations from foils (1) and (2) produced by the scattered portion of the beam, and the radiations sum to form a smooth background ($I = I_{1S} + I_{2S}$). Above this background the interference fringes from the unscattered beam are clearly visible. The effects of unscattered beam divergences $\sigma_U = 0$, 0.2 and 0.4 mrad are shown in Figure 8. The results indicate that the expected divergence, $\sigma = 200$ micro radians, can be readily determined with such an interferometer.

## 4 EXPERIMENTAL CONSIDERATIONS

The experiment proposed is to use an ODR-OTR interferometer to measure the divergence of the ATF accelerator operating at 30 MeV, using experimental methods that have been successfully employed used by us in conventional OTRI. The charge per pulse (0.5nC) for the ATF accelerator is the same for beams with which we have had success in these types of measurements previously. The position of the interferometer will be placed such that the beam can be focused (x and y) to a waist condition at the site of the interferometer. Polarized interferometry can then be used to measure the x and y divergences of the beam.

## 5 CONCLUSIONS

We have shown using simulation codes that the measurement of the divergence of a low energy low emittance, non-space charged dominated electron beam can be made using a perforated foil ODR-OTR interferometer. With the proper proportion of holes to solid material and size of the perforations, interference fringes produced by ODR and OTR from unscattered electrons passing through the perforations of the first foil have been shown to be clearly visible above the incoherent background radiation from the scattered portion of the beam. A proof of principle experiment using the ATF 30 MeV accelerator is now being designed to validate these code predictions. This new technique could be very useful to measure the divergence of low energy, high quality beams such as the 10 MeV MIRFEL accelerator being constructed at the University of Maryland, as well as high energy, low emittance beams.

**ACKNOWLEDGEMENTS**   This work is supported in part by the U. S. Government, Department of Energy; STTR Grant No. DE-FG02-01ER86132.

**REFERENCES**

[1] R. B. Fiorito and D. W. Rule, "Optical Transition Radiation Beam Emittance Diagnostics", in AIP Conference Proceedings No. 319, R. Shafer editor, (1994).

[2] M. L. Ter-Mikelian, **High Energy Electromagnetic Processes in Condensed Media**, Wiley-Interscience, New York, NY (1972).

[3] R. B. Fiorito and D. W. Rule, Nuc. Instrum. and Meth. B, 173, 67-82 (2001).

[4] V. Verzilov, Phys. Lett. A, 273, 135-140 (2000).

[5] A. P. Potylitsin, Nuc. Instrum. and Meth. B, 145, 169 (1998).

[6] I. Ben-Zvi, (private communication).

**FIGURE CAPTIONS**

Figure 1. Schematic of an optical diffraction-transition radiation interferometer.

Figure 2. Schematic of a single conducting foil showing an incoming electron associated wave (-e), an image charge wave (+e) and backward and forward directed optical transition radiation.  Note that the electron and its image charge travel together within the conductor.

Figure 3.  Demonstration of Babinet's Principle applied to the fields of diffraction radiation from a disk and a hole calculated using code BEAM DR.  Solid line: TR field from a continuous conductor; dotted line: DR from a disk; dashed line: DR from a circular hole (complementary to the disk); dashed-dot line : sum of the disk and hole fields (superimposed on the solid line).

Figure 4.  Comparison of BEAM DR calculations of the intensities of 1) OTR from a continuous conducting surface (solid line); 2) ODR from an electron passing right at the edge of a conductor (dotted line); 3) ODR from an electron passing at $l = 0.75\gamma\lambda/2\pi$ into the conducting surface (dashed

line) and 4) ODR from an electron in vacuum passing at $l = -0.75\gamma\lambda/2\pi$ from the conducting surface (dot-dashed line).

Figure 5. BEAM DR calculations of the intensity of transition radiation observed on a flat plane at various distances from the source.

Figure 6. Portion of a grid of rectangular holes in a metal screen showing the passage of a single electron through one hole and the surrounding electron fields. The dotted rectangle depicts the boundary of a unit cell that the code BEAM DR uses to calculate the diffraction radiation for a particular electron position in the beam.

Figure 7. BEAM DR calculations of a) two times the intensity of ODR from foil (1), b) two times the intensity of OTR from foil (2) and c) the sum of the product of the fields from these foils, for unscattered electrons, as a function of observation angle.

Figure 8. Effect of beam divergence on scans of horizontally polarized interference fringes observed from a ODR-OTR interferometer; beam energy, $E = 30$ MeV, foil separation $L = 30$ mm, observation wavelength, $\lambda = 650$ nm, band pass, $\Delta\lambda/\lambda = 0.03$, divergence due to foil scattering, $\sigma_S = 10$ milliradians.

FIGURE 1

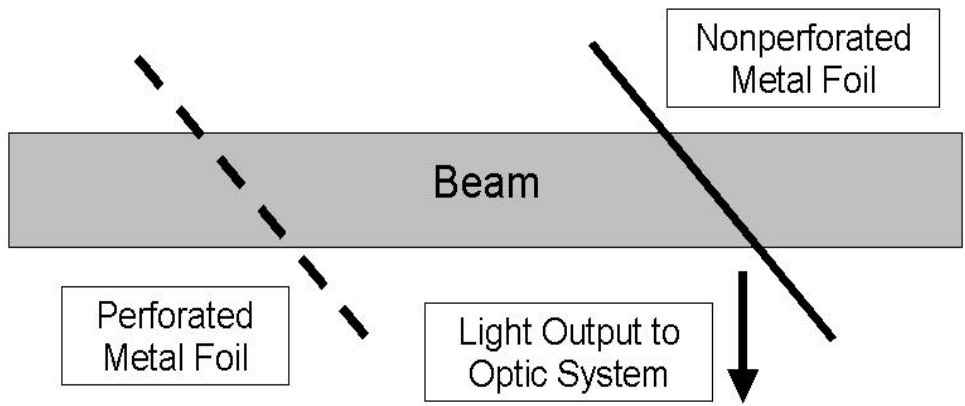

FIGURE 2

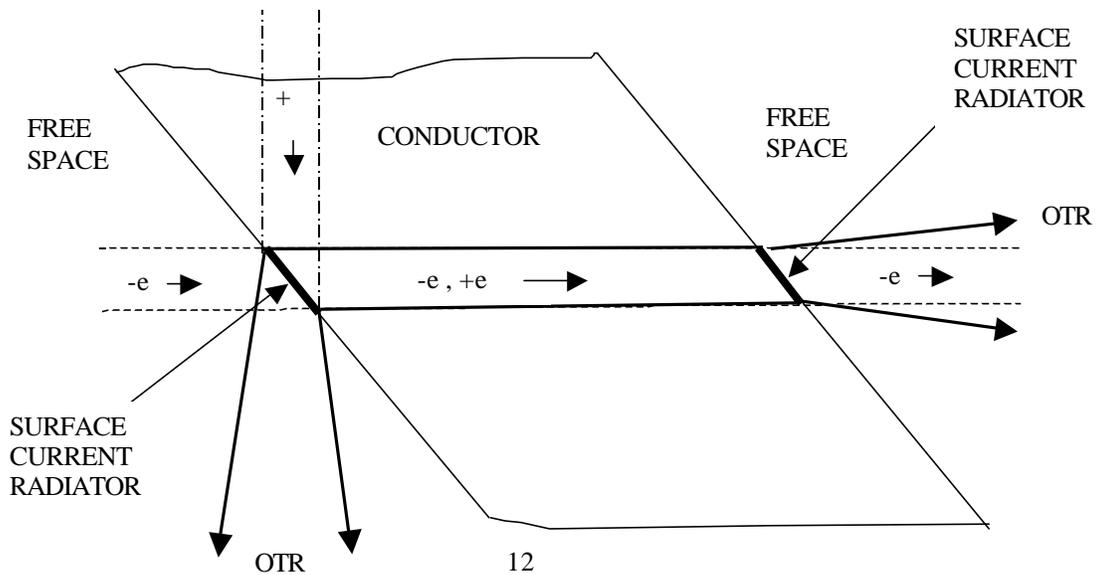

FIGURE 3

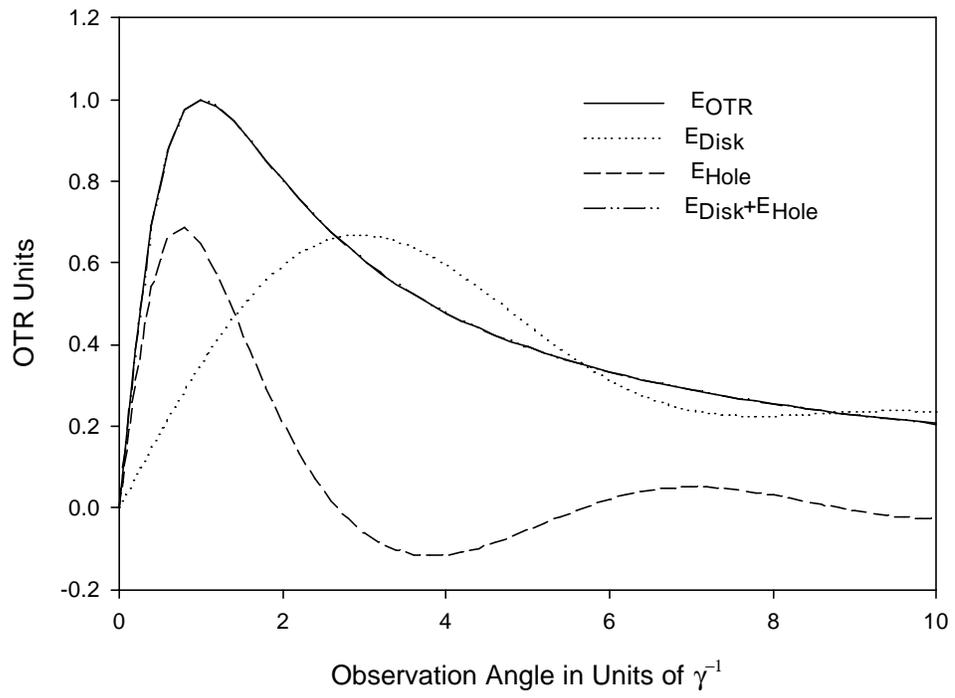

Observation Angle in Units of $\gamma^{-1}$

FIGURE 4

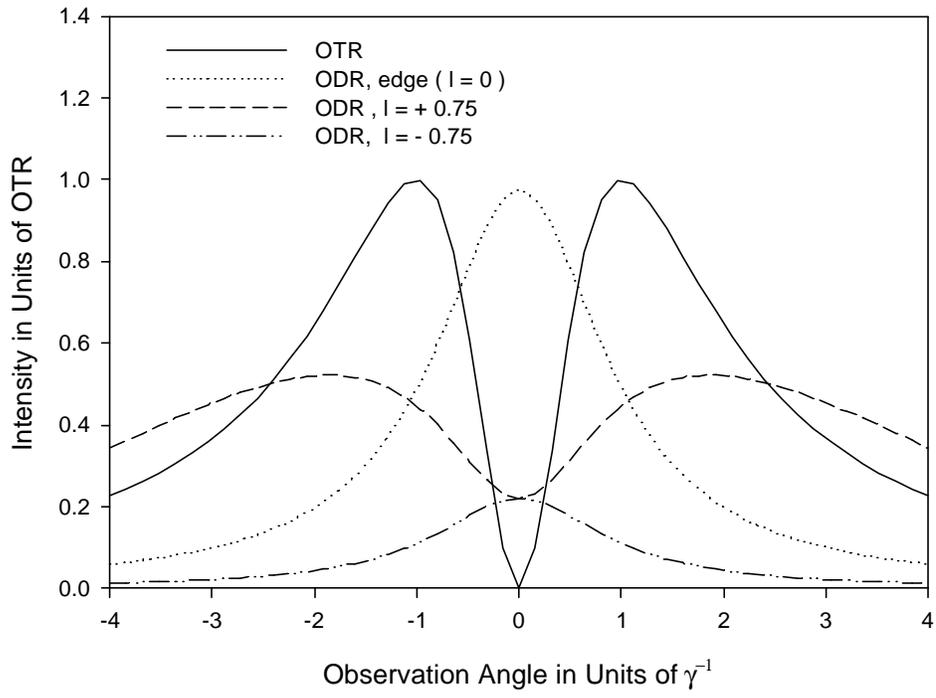

FIGURE 5

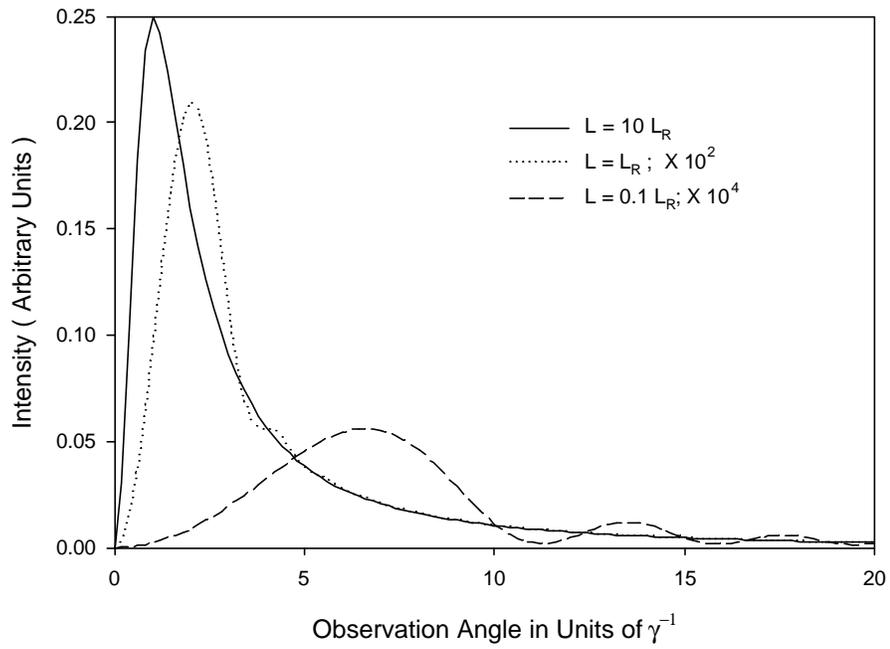

FIGURE 6

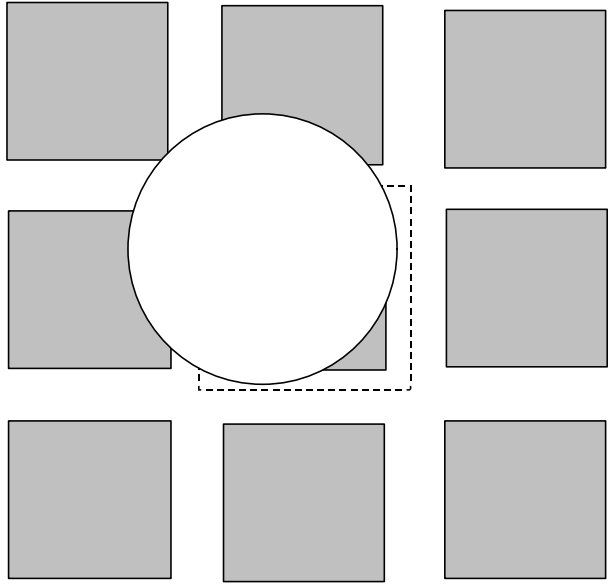

FIGURE 7

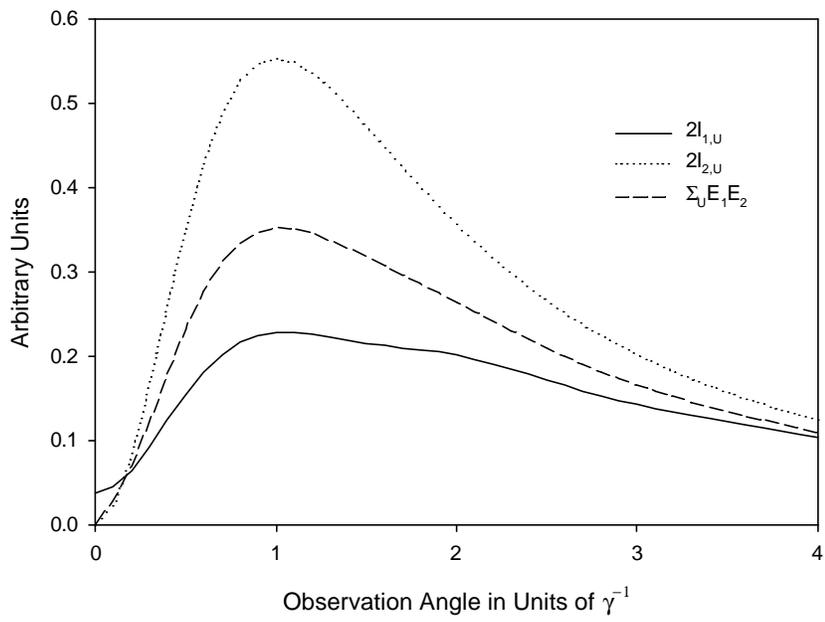

FIGURE 8

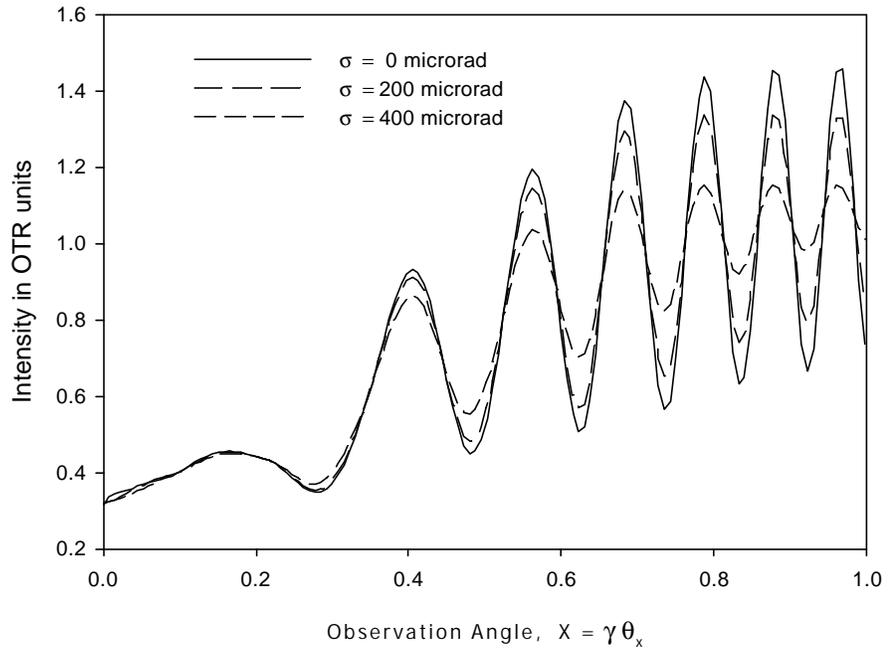